\documentstyle[12pt]{article}
\def\beginapjbib{\begingroup \section*{\large \bf References}
         \parskip=.5ex plus 1.0pt
         \def\bibitem{\par \noindent \hangindent\parindent
                \hangafter=1}}
\def\endapjbib{\par \endgroup}

\begin{document}
\begin{titlepage}
\pagestyle{empty}
\baselineskip=21pt
\begin{flushright}
UMN-TH-1206/93\\
July 1993\\
\end{flushright}
\vskip 0.5in

\begin{center}
{\large{\bf On the Galactic Evolution of $D$ and $^3He$}}
\end{center}
\begin{center}
\vskip 0.3in
{Elisabeth Vangioni-Flam$^1$, Keith A. Olive$^2$, and Nikos Prantzos$^1$
}\\

\vskip 0.25in
{\it

$^1${Institut d'Astrophysique de Paris, 98bis Boulevard Arago, 75014 Paris,
 France}\\

$^2${School of Physics and Astronomy,
University of Minnesota, Minneapolis, MN 55455, USA}\\

%$^3${Service d'Astrophysique, Centre d'Etudes de Saclay,
%91191 Gif sur Yvette, France}
}%\\
\vskip 0.5in
{\bf Abstract}
\end{center}
 \baselineskip=18pt

The determined abundances of primordial $^4He$ and $^7Li$ provide a basis
with which to test the standard model of big bang nucleosynthesis in
conjunction with
the other two light element isotopes $D$ and $^3He$, also produced
in the big bang. Overall,
 consistency in the standard big bang nucleosynthesis model
is best achieved for a baryon-to-photon ratio of
 typically $3 \times 10^{-10}$ for which
the primordial value of $D$ is five times greater than the
present observed abundance
and about three times greater than the pre-solar value.
We consider various models for the chemical evolution of the
 Galaxy to test the feasibility
for the destruction of D without the overproduction of $^3He$
 and overall metallicity.
Models which are capable of achieving this goal include ones
with a star formation rate
proportional to the gas mass fraction or an exponentially
 decreasing star formation rate.
We discuss the effect of parameters that govern the initial
 mass function and of
surviving fractions of $^3He$ in stars between one and three solar masses.
\end{titlepage}
%\newpage
\baselineskip=18pt
\def\d{\hbox{$D~$}}
\def\he#1{\hbox{${}^{#1}He$}}
\def\li#1{\hbox{${}^{#1}Li$}}
\def\be#1{\hbox{${}^{#1}Be$}}
\def\b#1{\hbox{${}^{#1}B$}}
\def\beq{\begin{equation}}
\def\eeq{\end{equation}}

{\newcommand{\la}{~\mbox{\raisebox{-.6ex}{$\stackrel{<}{\sim}$}}~}
{\newcommand{\ga}{~\mbox{\raisebox{-.6ex}{$\stackrel{>}{\sim}$}}~}

\section{Introduction}

As part of the foundation of our understanding of the early Universe,
the need to test
and scrutinize the standard model of big bang nucleosynthesis (SBBN)
 is essential.
Fortunately, SBBN is a testable theory.  It predicts the primordial
 abundances of
the light elements \d, \he3, \he4 and \li7 (see e.g. Walker et al. 1991).
 Unfortunately, one can not directly measure
the {\em primordial} abundances of these elements. With a certain
degree of confidence however, the abundances of \he4 and \li7 may be
extracted from observations. In the case of \he4, systematic observations
of low metallicity, extragalactic HII regions (see e.g. Pagel et al. 1992)
have led to
a fairly well determined value for primordial \he4 (Olive, Steigman and
 Walker 1991; Pagel et
al. 1992).  In the case of \li7, the observation of a well defined abundance
in old
halo dwarf stars (Spite \& Spite 1982a,b 1986; Hobbs and Duncan 1987;
Rebolo, Molaro and Beckman 1988), seemingly independent
of temperature (for $T \ga 5500$ K) and metallicity
(for $[Fe/H] < -1.3$) is generally regarded as a good indicator for the
primordial abundance.
Though there remains the problem of lithium depletion in stars, standard
 stellar models
(see e.g. Deliyannis, Demarque, \& Kawaler 1990) support the connection
 between the
observed abundances in halo dwarfs and the primordial value.  The case
for \d and \he3
is somewhat more complicated.

Deuterium is destroyed in stars. Despite the fact that all observed deuterium
 is primordial,
the primordial abundance cannot be determined from observations. Nevertheless
a useful upper
limit to the baryon-to-photon ratio, $\eta$, is established from the lower
limit
to the deuterium abundance (Gott et al. 1974).  A comparison between the
predictions
of the SBBN model and observed solar and interstellar values of deuterium
 must be made
in conjunction with models of galactic chemical evolution (Audouze and
Tinsley 1974).
The problem concerning \he3 is even more difficult.  Not only is primordial
 \he3
destroyed in stars but it is very likely that low mass stars are net
 producers of \he3.
Thus the comparison between theory and observations is complicated not
only by our
lack of understanding regarding chemical evolution but also by the
uncertainties in
production of \he3  in stars. Once again a useful lower limit to $\eta$
is obtained
by assuming that at least some of the \he3 in stars survives
(Yang et al. 1984).

SBBN has also been indirectly confronted by recent observations
 of $Be$ and $B$
(Rebolo et al.
1988; Ryan et al. 1990; Gilmore, Edvardsson \& Nissen 1992;
Ryan et al. 1992;
Gilmore et al. 1992;
 Duncan, Lambert \& Lemke 1992; Rebolo et al. 1993). In the big bang, these
elements are produced at abundance levels which are orders of magnitude below
that of the observations (Thomas et al. 1993; Delbourgo-Salvador and
 Vangioni-Flam 1993).
Though it is highly
likely that that both $Be$ and $B$ are produced by cosmic-ray spallation
 (Reeves et al. 1970; Meneguzzi et al.
1971) some $^7Li$ is also partly produced by galactic cosmic-ray
 nucleosynthesis
providing a potential constraint on big bang nucleosynthesis
(Steigman \& Walker 1992;
Olive \& Schramm 1992). The combination of a double source for
\li7 and the low value
for the abundance in the \li7 plateau all lead to a small
 value for the primordial
\li7 abundance. A higher primordial \li7 abundance can be
 tolerated in conjunction
with stellar models which deplete \li7 (see e.g. Pinsonneault,
Deliyannis,  \& Demarque 1992),
however, these models are constrained (Steigman et al. 1993) by
recent observations
of \li6 in a halo dwarf (Smith, Lambert \& Nissen 1993).

The results of SBBN with regard to constraints placed on what
 is perhaps the single remaining
parameter of SBBN, the baryon-to-photon ratio, $\eta$, can be
 summarized as follows
(Walker et al. 1991):
Extrapolations to zero metallicity of the helium abundance in
extragalactic HII regions
leads to a best value for the primordial helium mass fraction
(Olive et al. 1991;
 Pagel et al. 1992),
\begin{equation}
Y_p \simeq 0.23 \pm 0.01  \qquad  \eta_{10} \la 4
\end{equation}
where $\eta_{10} = \eta \times 10^{10}$. Note that only the upper
limit from \he4 is used
and one must be aware that relaxing the upper bound to 0.245
 relaxes the bound on
$\eta_{10}$ to $\sim 6$.
The best estimate for the pre-solar deuterium abundance
 (Geiss 1993) gives the upper bound on $\eta$
\begin{equation}
(D/H)_\odot \simeq (2.6 \pm 1.0) \times 10^{-5}  \qquad \eta_{10} \la 7
\end{equation}
In this case, it is only the lower bound on $D/H$ which is useful.
By assuming that at least 25\% of the primordial \he3 survives
 stellar processing
the upper limit on $(\d + \he3)/H$ gives a lower bound to $\eta$,
\begin{equation}
(D + ^3He)/H < 10^{-4}  \qquad  \eta_{10} > 2.8
\end{equation}
Finally with regard to \li7, if we assume neither \li7 depletion,
 nor a contribution from
cosmic-ray nucleosynthesis (these two effects work in opposite directions),
the mean \li7 abundance in the plateau of halo dwarfs is
\begin{equation}
^7Li/H = (1.2 \pm 0.1) \times 10^{-10}  \qquad  2.1 < \eta_{10} < 3.4
\end{equation}
When all of the light elements produced in the big bang are
 considered, consistency is achieved
when $\eta$ is in the range,
\begin{equation}
2.8 < \eta_{10} < 3.4
\end{equation}
In our forthcoming discussion concerning the destruction of \d and \he3,
we will for the most part keep $\eta_{10}$ fixed at the value
$\eta_{10} = 3$.  At this value of $\eta$, the primordial values of
$D/H$ and $^3He/H$ are approximately, $7.5 \times 10^{-5}$ and
$1.5 \times 10^{-5}$ respectively.

Larger values of $\eta_{10}$, such as $\eta_{10} \la 4$,
are possible given the
uncertainty in some of the nuclear reaction rates involving
\li7 (Walker et al. 1991).  With a larger value of $\eta_{10}$,
 smaller initial
values of \d and \he3 are required.  For example, at $\eta_{10} = 4$,
$(D/H)_p = 4.7 \times 10^{-5}$ and ($^3He/H)_p = 1.3 \times 10^{-5}$.
With this value of $D/H$, less deuterium destruction is required,
thus relaxing
this constraint on chemical evolution models.  As has been shown
by Steigman and Tosi (1992)
there are satisfactory models with higher values of $\eta$.

Given the SBBN as a presumption, we plan to study the evolution of \d and \he3.
With the above value of $D/H$, \d evolution is a necessity.
The abundances of \d and \he3 have been reviewed recently in Geiss (1993).
The pre-solar \d abundance, recall, is not measured directly.
Instead, a comparison is made between the \he3 abundance in carbonaceous
chondrites (in the noble gas component of meteorites which are unaffected by
solar deuterium burning) whose values are low and the higher \he3 abundances
measured in  gas-rich meteorites, the lunar soil and solar wind.
The former is representative of the true pre-solar \he3 abundance, while the
latter represents the sum of pre-solar (\d + \he3).
Amazingly, the pre-solar abundances of these isotopes has remained
 quite stable.
Measurements of \he3/\he4 in the solar wind with the ISEE-3 satellite
(Coplan et al. 1984)
show real fluctuations in the \he3/\he4 ratio.  Though the average
 value is the same
as the older measurements, there is most certainly a greater
uncertainty associated
with the pre-solar (\d + \he3) value.  Geiss has also
 increased the uncertainty in the
pre-solar \he3 abundance (in carbonaceous chondrites)
due to effects of fractionization.
In what follows, we adopt Geiss' pre-solar values for \d and \he3
(uncertainties are given at the 2 $\sigma$ level):
\begin{eqnarray}
((D + ^3He)/H)_\odot = (4.1 \pm 1.0) \times 10^{-5} \label{dhe3} \\
 (^3He/H)_\odot = (1.5 \pm 0.3) \times 10^{-5}  \\
(D/H)_\odot = (2.6 \pm 1.0) \times 10^{-5} \label{dsolar}
\end{eqnarray}

Deuterium is also measured in the ISM using Lyman absorption spectra
in nearby stars, thus we have a handle on the present day
\d abundance as well. Overall these measurements give a present
$D/H$ ratio of
$0.5$ to $2 \times 10^{-5}$ (see e.g. Vidal-Madjar 1991, Ferlet 1992).
A recent high precision
measurement of $D/H$ was made in the direction of Capella using
the HST Goddard High Resolution
Spectrograph (Linsky et al. 1992). The measured value
(and the one we will adopt for
the present day $D/H$ ratio) is
\begin{equation}
(D/H)_o = 1.5^{+.07}_{-.18} \times 10^{-5}
\label{do}
\end{equation}
Recent determinations (Bania, Rood \& Wilson 1993)
 of \he3 in the ISM yield values which range from $1.1$ to $4.5
\times 10^{-5}$, a domain which is still too broad to fully
 constrain models of chemical
evolution.

It seems timely to update, refine and generalize the analysis of the
 destruction of deuterium in the course of its galactic evolution since
the observed abundances of \d (as well as that of \he4)
have increased with respect to those considered by Vidal-Madjar and Gry (1984)
and Delbourgo-Salvador et al. (1985).
 Indeed, in the 1980's, it seemed necessary to invoke a small
surviving fraction
  of \d to obtain a primordial abundance in agreement with the
 big bang prediction. For that reason, Vangioni-Flam and Audouze (1988)
developed specific models aimed at destroying \d by a factor ten or even more.
 This problem, however, now seems less severe and a milder destruction factor
(4 to 5) is  sufficient.
  Moreover, the recent measurements of the \d abundance in the
local ISM (Eq.~\ref{do})
   and the protosolar ratio (Eq.~\ref{dsolar})
indicate that
 D has decreased since the birth of the sun.
We  also have better limits on the present gas mass fraction, $\sigma$,
  which is a key parameter in galactic evolutionary models.
 A typical estimate
 of the surface density of total matter at the solar circle from dynamical
  arguments is $54 \pm 8$~ $M_\odot/$pc$^2$ (Kuijken and Gilmore 1991).
This value combined with the gas surface density of 6 to 10
 $M_\odot/$pc$^2$, (Solomon 1993)
leads to an estimate of $\sigma$ between 0.1 and 0.2.
 Finally with regard to \he3, the situation remains vexing:
observational problems persist due to
   the dispersion of the measured interstellar abundances;
 and above all, there are considerable
  theoretical uncertainties on the production and destruction of \he3
   in low mass stars.

Our goal in this paper therefore, is to explore models of chemical
 evolution which have the possibility
of accounting for the destruction of deuterium from a primordial
value of $\sim 7.5 \times
10^{-5}$ to the pre-solar and present day values above, which in
 addition avoid
overproducing \he3. The same question was probed recently by
 Steigman and Tosi (1992).
Starting with a specific set of chemical evolution models
 (Tosi 1988), they constrained
the degree of deuterium destruction and hence the primordial
 deuterium abundance and $\eta$.
Here, we take a different approach. Given the level of
 uncertainty in models of chemical
evolution, which result from our lack of knowledge regarding
the initial mass function (IMF)
and perhaps more importantly the star formation rate (SFR), we
investigate to what extent plausible senarios of the chemical history of the
Galaxy can be reconciled with the factor of $\sim 5$ of total destruction
of \d.

\section{The Destruction of Deuterium}

Our goal in this section is to determine the conditions for which deuterium
may be destroyed by a factor of $\sim 2-4$
from its primordial value to
its pre-solar value and by a factor of $\sim 5$ to its present value.
The destruction of deuterium in connection with models of galactic
 chemical evolution
has been discussed somewhat extensively in the literature
(Audouze \& Tinsley 1974;
Ostriker \& Tinsley 1975; Audouze et al. 1976; Gry et al. 1984; Clayton 1985;
Delbourgo-Salvador et al. 1985; Vangioni-Flam and Audouze 1988;
 Steigman \& Tosi 1992).
There is no question that the degree of deuterium destruction is model
dependent.  Indeed the ratio $D/D_p$ (where $D_p$ is the
primordial abundance) depends on most aspects of a chemical
evolution model, which include the IMF, the SFR (Vangioni-Flam and
 Audouze 1988),
return fraction R (Ostriker and Tinsley 1975;
Clayton 1985), the infall rate (Audouze et al. 1976; Clayton 1985),
the composition of the infalling gas (Gry et al. 1984;
Delbourgo-Salvador et al. 1985),
and even computational approximations such as the often used
 instantaneous recycling
approximation (see below).

To calculate the abundance of deuterium as a function of time,
even in a simplified closed-box
model with no infall, one must still specify the IMF, SFR, and
 return fraction.
One can write down a simple analytic expression for $D/D_p$ which
involves only
the gas mass fraction, $\sigma$, and R (Ostriker and Tinsley 1975)
in the instantaneous
recycling approximation (IRA).  The gas  mass evolves as
\beq
\frac{d \sigma}{dt} = - (1-R) \psi(t)
\label{dmu}
\eeq
 where $\psi(t)$ is the SFR, and the return fraction is defined in terms of
the IMF, $\phi$, as
\beq
R = \int_{M_1}^{M_{sup}} (M-M_{rem}) \phi(M) dM
\label{R}
\eeq
In (\ref{R}), $M_1 \approx 0.85$ is the present main-sequence
 turn-off mass and $M_{sup}$ is the
upper mass limit for $\phi$.  $M_{rem} $ is the remnant mass:
$M_{rem} = M$ for $M < 0.5 M_\odot$,
$= 0.45  + 0.11 \ M/M_\odot$ for $0.5 < M/M_\odot < 9.0$
(Iben and Tutukov 1984)
and $= 1.5 M_\odot$ otherwise.
Correspondingly, the evolution of the deuterium mass fraction is
\beq
\frac{d(X_D\sigma)}{dt} = - X_D \psi(t)
\eeq
which can be combined with (\ref{dmu}) and easily solved
\beq
\frac{D}{D_p} = \sigma^{R/(1-R)}
\label{D}
\eeq
If we take $\sigma_o \approx 0.1$ as a representative
 present-day value, then
simple (power-law) IMF's  taken with $M_{rem}$ from
 above give $R \approx 0.2$
which in turn yields $(D/D_p)_o \sim 1/2$. A total deuterium
 destruction factor
of 2-3 is common in many models (Audouze and Tinsley 1975;
 Steigman and Tosi 1992).
However from the simple expression above (\ref{D}), for models
with a rapidly decreasing
SFR, the resulting IMF (as determined from the present-day mass
 function) in general
yields a larger value for R. When $R \sim 0.5$, deuterium will
 be reduced by a factor
of 10.

Infall introduces  another parameter which affects the deuterium abundance.
The degree of deuterium destruction depends on the composition of
the infalling
gas.  For a primordial composition, the total amount of destruction
is limited and
in many cases one finds a rise in the deuterium abundance from the
pre-solar value
to its present value (Gry et al. 1984; Steigman and Tosi 1992), which appears
to be in contradiction with the data.
In the  models considered by Steigman and Tosi (1992) the net
 destruction (primordial to present)
of deuterium was typically no larger than a factor of 2.
Thus better determinations of both the pre-solar and ISM
values of $D/H$ can be a valuable tool in limiting the amount of
 infalling gas.
If instead, the infalling gas composition is that of processed
 material, the $D/D_p$
ratio can be much smaller, $D/D_p \sim (1/10) - (1/40)$
(Gry et al. 1984; Delbourgo-Salvador et al.
1985).  Because of the apparent decrease in $D/H$ with time,
 and for the purpose of
simplicity as well as the lack of observational evidence,
we will not include infall in the subsequent discussion.
We simply note that the assumption of substantial (i.e. non negligible)
infall with primordial composition in the disk during
the last $\sim$5 Gyr may lead to an increase in the \d abundance and
to large (\d + \he3)$/H$ values and as such would be
inconsistent with the observations.

Depending on the specific model, the use of the IRA may also
affect the degree to which
\d is destroyed. For example, in Fig.~1, we show the evolution
of \d as a function
of time  with and without the use of the IRA.  In Fig.~1a, we have chosen
a single slope IMF (see below for a more complete description of these models)
and
a SFR, $\psi(t)$, which is proportional to the gas mass fraction, $\sigma$.
As one can see, the effect of the IRA on the destruction of \d is
reasonably small.
In Fig.~1b, we have chosen the Scalo (1986) IMF and
an exponentially decreasing SFR with a time constant of 3 Gyr.
This is a rather extreme case, leading to a current SFR much lower than
the past average SFR (see below) and much lower than observations of the
current SFR in the solar neighborhood (indicating that
the SFR is $\sim$3-5 M$_{\odot}$ pc$^{-2}$ Gyr$^{-1}$,
for a surface gas density of $\sim$10  M$_{\odot}$ pc$^{-2}$).
We adopt it as an extreme example of a large \d depletion.
It can be seen from  Fig.~1b that in this case
the effect of the IRA is significantly more important.

It is interesting to analyze the effect of removing the IRA. It turns out that
\d destruction is correlated to the evolution of
the gas fraction $\sigma(t)$.
In Fig.~2a we see that the evolution of $\sigma$ is approximately the same
with and without the IRA in the case where
$\psi(t) = \nu \sigma$. This can be explained in the following way:
since the gas evolves relatively slowly, the amount of (\d poor)
ejecta is small compared to $\sigma(t)$ at any time.
\d destruction depends then little on the assumption of an
instantaneous recycling.
On the other hand, when $\psi(t) = exp(-t_{Gyr}/3)$, $\sigma$ declines
very sharply early on and
in the IRA, the amount of matter ejected becomes significant with
respect to $\sigma$. Indeed, in Fig.~2b one finds significantly more gas
at intermediate times ($t \simeq$a few Gyr) when the IRA is made, which is
largely composed of the, instantly returned, \d poor ejecta; consequently, the
IRA leads to a larger \d depletion at that period. At late times, however, the
IRA and non-IRA give similar amounts of gas (Fig 2b); indeed,
for such late times all of the stars that can return an important fraction of
their mass, have enough time to do it. But the IRA stars, being created in
small numbers at late times (small SFR) cannot considerably dilute the \d
abundance of  the gas with their ejecta;
on the contrary, the large number of  long
lived stars that were created early on in the non-IRA case, return a large
(with respect to the late gas) \d free amount of matter, considerably diluting
the \d abundance at late times. Thus, the IRA approximation in the case
of a rapidly decreasing SFR overestimates the \d depletion at early times and
underestimates it at late times.
The difference in the behaviour of \d (and \he3) in the
case of the IRA and non-IRA
calculations is much more apparent when the results are plotted as a
function of the gas fraction; this is done in Fig. 3, which nicely illustrates
the previous discussion.

The impact of the SFR on the degree of
deuterium astration was studied extensively by Vangioni-Flam
and Audouze (1988, VFA). We summarize that work here as it will
 serve as a basis to our present work and its general philosophy remains
   pertinent.
  In VFA, two kinds of solutions had been proposed to astrate \d efficiently:

a) - A high SFR in the early galaxy, with a normal IMF( model II in VFA).
   In this case, the \d poor
  gas is ejected, on average, after a long delay, due to the large number
  of low mass stars. The overproduction of \he3 is
avoided under the reasonable
 assumption that
about 30\% of the original \d + \he3 survives in the form of \he3
in stars between   1 and 3 M$_\odot$.
This model had been subsequently discarded after it was put to the test
 using the G-dwarf metallicity distribution (Francois, Vangioni-Flam and
 Audouze 1989);

b) - A modified IMF favoring massive stars (model IV in VFA).
 In this case, the \d-free
   gas is released almost instantaneously by massive stars, and the IMF
   must be adjusted to avoid excess production of $^{16}O$ and metals.
   It is worth mentioning that model I (in VFA) with a SFR proportional to
   the gas fraction destroys  \d by a factor of 3.3, which is not too
  far off from the new required value. Perhaps the time has come to reconsider
   this kind of simple model.

 Exponentially decreasing SFRs have also been explored
   by Olive, Thielemann \& Truran (1987, OTT) in the
 framework of the IRA. A constant IMF derived
  from the present-day mass function (PDMF)
 was used together with its associated SFR (Scalo 1986).
   Apparently, this is a good candidate because this special
  combination of SFR and IMF offers an efficient way to
  lower the \d abundance and a possible solution to the G-dwarf
 problem (Olive 1986) at the same time.
   There are some similarities between these models and model IV of VFA.
   The overproduction of $O$ and metals is avoided at the expense of imposing
 a cutoff at the high mass end of the IMF.
  We will also consider a substitution of  Scalo's IMF by a power
  law one in order to try to avoid as much as possible an excess of
metals produced by massive    stars.

\section{\he3 Production and Destruction}
Even more complicated than the history of deuterium,
 is that of \he3.  Not only
does the abundance of \he3, as a function of time,
depend on standard galactic evolution parameters such as the IMF, SFR, etc.,
but also on the production of \he3 inside a star and its return to the ISM.
While there is little debate that in more massive stars ($M > 5-8 M_\odot$)
\he3 is efficiently destroyed, in low mass stars ($1M_\odot < M < 2M_\odot$)
\he3 is perhaps produced, some of which will be returned to the ISM.
It is precisely because of the likelihood that not all of the primordial \he3
is destroyed in stars, that the measurement of pre-solar $D +~^3He$ can
be used to set a lower limit on $\eta$.

It was noted in Rood, Steigman and Tinsley (1976),
that by requiring that \he3 not be
overproduced, a lower limit to $\eta$ could be set.
The limit disappears if \d and
\he3 destruction is complete.
The argument yielding a lower limit to $\eta$ based on pre-solar $D +~^3He$
was first given in Yang et al. (1984). The argument runs as follows:
First, during pre-main-sequence collapse, essentially all of the primordial
\d is converted into \he3.  The pre-main-sequence produced and primordial
\he3 will survive in those zones of stars in which the
 temperature is low, $T \la
7 \times 10^6$ K. In these zones \he3 may even be produced by $p-p$ burning.
At higher temperatures, (up to $10^8$ K), \he3 is burned to \he4.
If $g_3$ is the fraction of \he3 that survives stellar processing,
then the \he3 abundance at a time $t$ is at least
\beq
\left( {^3He \over H} \right)_t \ge g_3 \left( \frac{D +~^3He}{H} \right)_p -
g_3 \left( {D \over H} \right)_t
\label{yl}
\eeq
The inequality comes about by neglecting any net production of \he3.
Of course, Eq. (\ref{yl}) can be rewritten as a upper
 limit on (\d + \he3)$/H$ in terms
of the observed pre-solar abundances ($t = \odot$) and $g_3$.

The models of Iben (1967) and Rood (1972) indicate that
 low mass stars, $M < 2 M_\odot$
are net producers of \he3.  For stars with mass $M < 8 M_\odot$,
 Iben and Truran (1978)
have estimated the final surface abundance of \he3,
\beq
(^3He/H)_f = 1.8 \times 10^{-4}\left({M_\odot \over M}\right)^2 +
0.7\left[(D+~^3He)/H\right]_i
\label{it}
\eeq
indicating that $g_3 > 0.7$, notwithstanding the uncertainties
 involved in determining (\ref{it}).
For more massive stars, Dearborn, Schramm \& Steigman (1986)
 have estimated $g_3$ for a
variety of metallicities and \he4 abundances. Overall, they
find $g_3$ in the range 0.1 to 0.5
for stars with $M > 8 M_\odot$.

For the purposes of obtaining a lower limit to $\eta$
based on $(D+~^3He)/H$, it is necessary
to estimate a lower limit to $g_3$.  Without this lower
 limit, heavy destruction of deuterium
and \he3 would allow for very low values of $\eta$
 (Olive et al., 1981).  A troubling aspect
of this argument has always been the lack of observational support.
Recently however, Ostriker and Schramm (1993) have argued
 that on the basis of observations
by Hartoog (1979) of \he3 in horizontal branch stars, a lower
 limit of $g_3 > 0.3$ could
be inferred.  Also, the high value of \he3$/H$ observed in a planetary
nebula by Rood, Bania \& Wilson (1992) seem to support
 the idea that \he3 is in fact not completely
destroyed and may be produced.

In our subsequent calculations, we will use the values
of $g_3$ as given in Dearborn et al. (1986).
However, for the mass range 1 to 3 $M_\odot$, which we
 find crucial for determining the pre-solar
deuterium plus \he3 abundance, we will consider several
 possibilities. We will refer to
$g_3$ as a set of three values corresponding to estimates
 of $g_3$ at $(1,2,3)M_\odot$
respectively. For comparison, Dearborn et al. (1986) used $g_3 = 1$ for
$M < 3M_\odot$, and
 $g_3 = 0.7$ for $3 < M/M_\odot < 8$. Delbourgo-Salvador et al. (1985) used
$g_3 =0.7$ for $M < 2M_\odot$, $g_3 = 0.25$
for $2 < M/M_\odot < 5$, and $g_3 = 0$ for
$M > 5 M_\odot$.
 For a more complete discussion on estimates of $g_3$ we refer the reader to
Yang et al. (1984) and Dearborn et al. (1986).

It is important to note that the \he 3 survival factors
considered previously and here
are all significantly lower than the Iben and Truran value
 at $1 M_\odot$, $g_3 = (1.8 \times
10^{-4})/[(D+~^3He)/H]_i +0.7$, which even for the relatively high
value $[(D+~^3He)/H]_i = 9 \times 10^{-5}$, gives $g_3 = 2.7$.
  As we will see,
such a large survival fraction
will prove to be irreconcilable with the pre-solar
abundance determination.  Unfortunately, we can offer no
solution as to why $g_3$ should
be lower other than the constraints imposed by the pre-solar $D+~^3He$ data.

In addition to its sensitivity to $g_3$, the \he3 abundance as a function
of time is also quite dependent on the parameters of the galactic chemical
evolution model (as for \d).
In the IRA, an analogous expression to (\ref{D}) was
 derived in Olive et al. (1990)
\beq
\frac{(D+~^3He)}{(D+~^3He)_p} = \left(\frac{D}{D_p}\right)^{g_3 - 1}
\eeq
so that the model dependence of $D/D_p$ feeds into $(D+~^3He)$.

{}From Fig.~1, we see how $(D+~^3He)$ is affected by the IRA.
In both cases (1a and 1b) the sum of (\d + \he3) is correlated,
 although not in
a straightforward way, to the fate of \d.
In Fig.~1a  there is more \d with the IRA in the end than without
 the IRA (as explained
in the previous section).
Since \d is not severely depleted in that case (it constitutes $\simeq$half
of the (\d + \he3) amount), the final (\d + \he3) abundance
 is also larger in the IRA case.
On the other hand, in Fig. 1b, \d suffers a severe depletion early on, and
its weight in the final (\d + \he3) sum is small. What matters then is the
final \he3 amount. In the non-IRA case a lot of \he3 at late times comes from
the early created long-lived stars that were enriched in \d; \he3 is then
quite abundant, as is the sum (\d + \he3). In the IRA case most of \he3 comes
from more recently created stars, that are \d poor; its final
 abundance is then smaller
than in the IRA case, and the same is true for the sum (\d + \he3).
We see then that the IRA has an opposite effect on the (\d + \he3) sum
in those two cases.

In any case, the abundance of (\d + \he3) compared to
the pre-solar value (\ref{dhe3})
will turn out to be among the toughest challenges to overcome.
In short, it is not the destruction of deuterium that is problematic, but
rather the overproduction of \he3.

\section{Galactic Evolution Models}

\subsection{Constraints}

The point now is clear: as shown over the last ten years,
the destruction of deuterium is highly model dependent.
  Our objective is to reduce the  SBBN
     \d abundance by a factor $\sim$5 over the galactic lifetime, reproducing
   at the same time the solar oxygen abundance and global metallicity $Z$,
for a galactic age of $\sim$14 Gyr and a current gas
fraction 0.1 $< \sigma <$
0.2.
 $[O/Fe]$ vs $[Fe/H]$ and $[Fe/H]$ vs time are also consistent
 with observations
 in all of the models proposed in this
   study and before (e.g Francois et al. 1989, Vangioni-Flam,
Prantzos \& Chauveau 1993).

  Previous work has also shown that the adopted formalism
complemented by specific
   treatment of cosmic ray spallation accounts for the
 evolution of Be and B as well
( Prantzos et al. 1993). In this context, the evolution
of $^9Be/H$ vs. $[Fe/H]$ and $^{11}B/H$ vs. $[Fe/H]$
are largely independent of the IMF and SFR (Prantzos et al.
 1993; Olive et al. 1993).
Consequently, we find similar results as before.

Finally as we are considering models with a time-varying SFR,
it is worthwhile to note
the constraints on the history of the SFR. For this purpose
 it is useful to define
the relative birthrate, $b(t) = \psi(t)/\langle \psi \rangle$
where the average SFR is defined
by
\beq
\langle \psi \rangle = {1 \over T_o}\int_0^{T_o} \psi(t)dt
\eeq
and $T_o$ is the age of the Galaxy.  For a constant SFR $b=1$,
while for an exponentially
decreasing SFR, $b(T_o) = (T_o/\tau) ( e^{T_o/\tau} - 1)^{-1}$.
 For $T_o$ = 14 Gyr
and $\tau$ = 3, 5, and 10 Gyr, $b(T_o)$ = 0.044, 0.18, and 0.46
 respectively. When
$\psi = \nu \sigma$, with $\nu = 0.25$, $b(T_o)$ = 0.33.
Limits on $b(T_o)$ have been reviewed by Scalo (1986): among the strongest
limits is $b(T_o) > 0.4$ derived from stellar age distributions. Though there
is a certain degree of uncertainty in this bound (see eg. Tinsley 1977,
 Twarog 1980),
it does give us an indication that $\tau = 3$ Gyr should be viewed
 as an extreme value.
However, in bimodal models of star formation (Larson, 1986; VFA),
 where only one component
(the massive end) has a rapidly decreasing SFR, the value
 $\tau = 3$ may still be plausible
and satisfy the age constraints.

\subsection{ Models and yields}

The yields of Woosley (1993) are adopted in this work. They are not
very different from other recent works (Arnett 1991;
Thieleman et al. 1993;  Weaver and Woosley 1993)
at least as far as oxygen
is concerned (see Prantzos 1993 for a comparison and implications).
The oxygen yield is determined within a factor of
$\simeq 2$ due to uncertainties in the $^{12}C(\alpha,\gamma)^{16}O$
 reaction rate
and in the treatment of convection in massive stars.
(For a discussion see OTT, and Weaver and Woosley 1993).
We performed a limited check of the impact of those yields by running a
standard model of chemical evolution (closed box, power-law IMF between
0.4 and 100 M$_{\odot}$, SFR $\propto \sigma$), the results of which appear
in Table 1.

The iron yield from core collapse supernovae (originating from massive stars)
 is
unfortunately still uncertain except for the $20 M_\odot$ case ($\sim$0.07
M$_{\odot}$ of Fe is produced, after the interpretation of the light-curve of
SN1987a).  The adopted yields of Woosley have been adjusted to get reasonable
values of $[O/Fe]$ ($\sim$0.5) in low metallicity stars, as was done
in our previous work.  Also uncertain is the past rate of
supernovae of type Ia.
We have chosen a constant SNIa rate of 0.2 per century, with each ejecting
0.6 M$_\odot$ of $Fe$.  The present ratio SNIa/SNII $\simeq 0.1$ is
required to be reproduced and
this simple procedure is generally sufficient to fit the steepening of the
$[O/Fe]$ vs. $[Fe/H]$ curve beyond $[Fe/H] \simeq -1$.

Finally, the destruction of \he3 in low mass stars
 ($1 \le M/M_\odot \le 3$) which is
poorly known (see section 3) has been treated as a
 free parameter within reasonable
limits.

Two sets of models have been selected, differing by their SFR. In model $I$,
the SFR $\psi(t)$, is assumed to be proportional to the gas mass fraction.
The constant of proportionality, or astration rate is $\nu = 0.25$ Gyr$^{-1}$.
The IMF is parametrized as $\phi(M) \propto M^{-(1+x)}$ with the normalization
\beq
\int_{M_{inf}}^{M_{sup}} M \phi(M) dM = 1
\eeq
where the slope $x$, and the mass limits, $M_{inf}$ and $M_{sup}$ are taken to
be variable.
In model $I_a$, a single slope IMF is considered between
the limits $0.4 < M/M_\odot < 100$.
We have also tested these models with various choices of the
\he3 survival fraction, $g_3$.
In model $I_b$, the Tinsley (1980) IMF is used between
$0.1 < M/M_\odot < 100$.
This traditional model is worth rehabilitating in the
 present context for \d destruction.

Model $II$, inspired by Scalo(1986), Larson (1986) and OTT,
 features an exponentially decreasing
SFR, $\psi(t) \propto e^{-t/\tau}$.
In Larson's model (a model of bimodal star formation)
the steeply decreasing SFR for high mass stars leads to a
 large density of low-luminosity
white dwarfs.  The price to pay in such a model however,
 is a somewhat unusually low
value for $M_{sup}$ to avoid the overproduction of metals, primarily oxygen.
In Larson's model, $M_{sup} = 16 M_\odot$. Using an
 exponentially decreasing (non-bimodal)
SFR, as we have considered here, OTT derived limits
 on $M_{sup}$ as a function of $\tau$.
Arguing further that if the yields of massive stars were well understood,
the abundance patterns such as $[C/Fe] \approx 0$, $[O/Fe] \approx 0.5$,
and $[(Ne + Mg + Si)/Fe] \approx 0.5$ could place a strong limit
on $\tau$ and $b(t)$. For example the yields of
Arnett (1978) typically require the presence
of massive ($M \ga 40 M_\odot$) stars, leading to a limit $b(t) \ga 0.75$.
However the yields of Woosley and Weaver (1986) require only the presence of
$\sim 15 M_\odot$ stars
and the limit on $b(t)$ drops to $b(t) \ga 0.03$.
In model $II_a$, we considered several values
 of $\tau$ using the Scalo (1986) IMF.
In model $II_b$ the single slope IMF was again employed.
A summary of the different parameters used in the two sets of models is
shown in Table 2.

Figure 4 shows the cumulative metallicity distribution
of the models in the galactic disk phase  (for $[Fe/H] \ga -0.7$),
compared to observations; the agreement is satisfactory, (though in model
$I_a$ it is less so) i.e. the G-dwarf
problem is solved, since the disk starts with an initial
metallicity enrichment
from the previous (halo) phase.

\section{Results}

In Fig.~5a we show the evolution of $D/H$ as a function of time for the four
basic models considered. As one can see from Fig.~5a,
there is a great variability in the degree of
destruction of \d. With the exception of model $II_{a,a}$,
(chosen to be extreme), all
of the models give a perfectly adequate picture for the
time evolution of \d.
(The present value inferred from model $I_a$ is perhaps slightly high compared
to
the HST measured ISM abundance.)  Thus one of our primary
 goals is achieved:
Starting with a primordial abundance $D/H = 7.5 \times 10^{-5}$,
 we are in fact able
to obtain destruction factors of 3-5 to agree with the
pre-solar and present-day
measurements.

Models without an extremely large return fraction
($R\sim 0.40$ in model $I_a$ and
$R\sim 0.54$ in model $I_b$) and a ``reasonably
smooth" SFR ($\propto \sigma$)
reproduce the observations quite satisfactorily. This is also true for
model $II_b$ (also with $R \sim 0.40$)
which has a rapidly decreasing SFR ($\propto exp(-t_{Gyr}/3)$),
and depletes \d somewhat more than the other two.
On the other hand, models with a steeply decreasing
 SFR (models $II_{a,a}$ and $II_{a,b}$
with $R \sim$ 0.65 and 0.52 respectively), can be ``lethal" to \d.
 This is the case of models $II_{a,a}$
(SFR $\propto exp(-t_{Gyr}/3)$)
and $II_{a,b}$ (SFR $\propto exp(-t_{Gyr}/5)$),which
destroy \d by a factor of $\sim$100, as can be seen in Fig.~5b. Model
$II_{a,c}$ (with $R \sim 0.42$) has a slowly
 decreasing SFR ($\propto exp(-t_{Gyr}/10)$) and
 gives a quite acceptable fit to the \d observations
(Fig.~5b again).
Finally, Fig.~5c illustrates the role of the IMF
in the depletion of \d, for the case
of a power-law IMF with $x=1.7$ and a smooth SFR
(= 0.25 $ \sigma$), i.e. model $I_{a,e}$.
When $M_{inf} = 0.1 M_{\odot}$ ($R\sim 0.17$) the depletion of \d is small
(less than a factor of two); it becomes compatible with the
observations when $M_{inf} = 0.4 M_{\odot}$ (and $R \sim 0.40$).

As noted earlier, it is more difficult to satisfy (with any model) the
constraint imposed by the pre-solar $(D+~^3He)/H$ value. Figs.~6a, 6b and 6c
show the evolution of (\d + \he3) corresponding to the models presented
in Figs.~5a, 5b and 5c, respectively.

{}From all of the models in Fig.~5a (that reproduce the \d evolution well),
only one ($I_b$) can also satisfy the (\d + \he3) constraint when the \he3
survival fraction is $g_3 = (1.0,0.7,0.7)$.
In the other two cases ($I_{a,e}$ and $II_b$)
it was necessary to reduce $g_3$ in order
to bring about agreement with the data.

This difficulty in obtaining an acceptable pre-solar (\d + \he3) is not
shared by models with a steeply decreasing SFR, as can be
seen in Fig.~6b. Models $II_{a,a}$ and $II_{a,b}$
reproduce nicely the observed
(\d + \he3) value, even with the larger value of
 $g_3$ value for the survival fraction
of \he3. The reason is, of course, that they destroy
 so early (and efficiently)
their \d, that
even if a considerable fraction of it survives in the form of \he3, the sum
is still reasonably low. But, as seen in Fig.~5b, those two models destroy
too much \d, and as such should be excluded. On the
 contrary, model $II_{a,c}$
(reasonably reproducing the \d evolution)
needs again a lower $g_3$ value (0.5,0.3,0.3) in order to reproduce
marginally the pre-solar $(D +~^3He)/H$ value.
Finally, Fig.~6c illustrates the effect of the IMF on the
(\d + \he3) evolution. The low value of
$M_{inf} = 0.1 M_{\odot}$ leads to a pre-solar (\d + \he3) much larger than
observed. A larger
$M_{inf} = 0.4 M_{\odot}$ brings about a better agreement
 with the observations.

It now becomes straightforward to understand the excess
(if $g_3$ is not lowered)
\d + \he3 in models $I_a$ and $II_b$.
It all has to do with the distribution of stellar masses.
Too much mass in low mass
stars will release an excess of \he3. As we saw, this effect is displayed by
 lowering $M_{inf}$ from
$0.4 M_\odot$ to $0.1 M_\odot$ in model $I_{a,e}$.
In this case because of the single
slope IMF, when $M_{inf} = 0.1 M_\odot$ there is significantly
 more mass in very low mass
stars which do not evolve.  The result is clear.  Much less \d is destroyed,
and there is an excess of \d + \he3 as was shown in Figs.~5c and 6c.
Model $I_a$ does worse than $I_b$ despite the fact that
model $I_b$ goes down to lower masses because the Tinsley (1980)
 IMF begins to turn over (though less
sharply than does the Scalo (1986) IMF) and has a much
 flatter slope ($x = 0.25$) compared
to the $x=1.7$ slope in $I_a$. The shape of the IMF at low mass
is of critical importance in determining the late-time
 behavior of \d and \he3.
Similarly, $II_b$ does worse than $II_a$ with respect to \d + \he3.

Fig.~7 presents the evolution of \he3  for several of
 the models discussed in
this section, and helps understanding the results on (\d + \he3) presented
in Figs.~6a, 6b and 6c. The survival fraction of \he3 used
in models $II_{a,a}$ and $I_b$ leads naturally to a large
increase in the \he3
abundance. On the other hand, the small $g_3$ value adopted in models
$II_b$ and $I_{a,e}$ (as to lead to an acceptable pre-solar \d + \he3), leads
to a very slight increase in the abundance of \he3 during galactic
evolution.

The importance of the \he3 survival factor was already recognized by
Truran and Cameron (1971). They showed
that the assumption of \he3 survival in low mass stars
 stars had a profound effect
on the comparison to data at the time of the formation of the solar system.
The potential for excess \he3 was also discussed in Rood et al. (1976).
In Fig.~8, we
perform a more systematic investigation of this  effect,
lowering $g_3$ from the more conservative value used in model
$I_{a,a}$ to almost half
of it in model $I_{a,e}$. As expected, the effect is rather large,
 but only for the
lowest value is a (marginal)  agreement to observations obtained.
A renewed effort in better understanding
this parameter is certainly needed.

An overall summary of our results is displayed in Table 3.

\section{Conclusions}

We have shown that the degree of \d astration during galactic
evolution depends
crucially on the adopted
stellar IMF and the star formation rate. We found that
it is not very difficult to destroy \d in the context of relatively
standard models of galactic chemical evolution without infall.
Without knowing anything about the primordial infall
rate, it seems that this might be a good approximation, especially in light
of the recent
\d observations which indicate that the \d abundance has decreased
 since the formation of
the solar system.  Future observations will certainly help our
 understanding of this issue.

However, in spite of the ease in destroying \d, one should remain
cautious regarding
the evolution of the other elements, such as \he3 and $^{16}O$.
Among the most difficult
of the constraints to satisfy is the limit
$(D+~^3He)/H \la 5 \times 10^{-5}$ on the
\d and \he3 pre-solar abundance. In particular
 \he3 production in stars with mass $\sim
1 M_\odot$ should be minimal.  It would be very useful
 to study in detail the production
and destruction of this element in the evolution of low mass stars
including the Asymptotic Giant Branch phase, since it is
ultimately more constraining than \d.

Considering the new requirements for the destruction of \d,
our model $I$, with the SFR proportional
to the gas mass fraction, is a good candidate for $D(t)$ as
well as $Z(t)$ {\it if}
\he3 is not overproduced in low mass stars.  Model $II$, on
the other hand, can
significantly destroy \d without affecting \he3, but the
overproduction of $O$
requires cutting off the high mass end of the IMF. Thus our
 main conclusion is
that a destruction of \d by factor of $2-3$ at the time of the
formation of the solar system
and a factor $\sim 5$ today, can be achieved with relative ease
in a variety of models
without infall.  However, our understanding of the net
 production of \he3 in stars of about
one solar mass is crucial to determining the overall
 viability of these models.
Somehow, $g_3 \le 1$ for  $M \simeq 1 M_\odot$,
i.e. there should not be significant net \he3 production in such stars,
and thus  must be below the estimate of Iben and Truran (1978).

A potential cop-out to this predicament may yet be that the solar
abundance of \he3 is not
representative of the average ISM abundance at that time (Rood et al. 1976).
Indeed, at $t=T_o$, none of our models produce \he3 in excess
of the current limits
on the present-day ISM abundance as measured by
Bania et al. (1993). Perhaps the solar
system is anomalously low in \he3.

}
}
\newpage
%\vskip 1.0truecm
\noindent {\bf Acknowledgements}
\vskip 0.3truecm
 We would like to thank
Jean Audouze, Michel Cass\'{e}, David Schramm and Jim Truran
for very helpful discussions.
KAO would like to thank the Institut d'Astrophysique de Paris where
this work was started for their hospitality.
The work of KAO was supported in part by  DOE grant DE-AC02-83ER-40105.
and by a Presidential Young
Investigator Award. The work of NP and EV-F was supported in
 part by PICS $n^o$114,
``Origin and Evolution of the Light Elements", CNRS. We are
grateful to Yvette Oberto
for her help in the calculations.
\vskip 1in
\begin{table}[h]
\begin{center} {\sc Table 1.  Abundance comparison at the birth
 of the solar system
using the yields of Arnett (1991) and Woosley (1993) in the case
of model $I_a$}
\end{center}
\begin{center}
\begin{tabular}{lcc}                                           \hline \hline
                    & Yields from Woosley           & Yields from
Arnett \\ \hline
$Z/Z_\odot$                  & 1.4         &   1.4    \\
$O/O_\odot$                        &2          &    1.4   \\
$Fe/Fe_\odot$           & 0.85         &   0.75     \\ \hline
\end{tabular}

\end{center}
\end{table}
\newpage
\begin{table}[h]
\begin{center} {\sc Table 2.  The list of models explored in our study.}
\end{center}
\begin{center}
\begin{tabular}{lccccc}
\hline \hline
 model   & SFR & IMF &$M_{inf}$ & $M_{sup}$ & $g_3$ \\ \hline
$I_{a,a}$& $0.25 \sigma$  &   $M^{-2.7}$ & 0.4 & 100  & (1.0,0.7,0.7)   \\
$I_{a,b}$& " &  " & "  & " & (1.0,0.5,0.5) \\
$I_{a,c}$& " &  " & "  & " & (0.7,0.5,0.5) \\
$I_{a,d}$& " &  " & "  & " & (0.7,0.3,0.3) \\
$I_{a,e}$& " &  " & "  & " & (0.5,0.3,0.3) \\
$I_b$ & " &Tinsley&0.1& " & (1.0,0.7,0.7)   \\
$II_{a,a}$&$e^{-t/3}$&Scalo&0.1&20& "   \\
$II_{a,b}$&$e^{-t/5}$&"&"&"&" \\
$II_{a,c}$&$e^{-t/10}$&"&"&100&(0.5,0.3,0.3) \\
$II_b$&$e^{-t/3}$&$M^{-2.7}$&0.4&100&(0.5,0.3,0.3)     \\ \hline
\end{tabular}

\end{center}
\end{table}
\begin{table}[h]
\begin{center} {\sc Table 3.  Results of four basic models compared
with observations
in the solar system as well as in the ISM for \d and $\sigma$.}
\end{center}
\begin{center}
\begin{tabular}{lccccc}
\hline \hline
             & Observations & Model $I_{a,e}$ & Model $I_b$ &
Model $II_{a,a}$ & Model $II_b$ \\ \hline
$\sigma$     & 0.1 to 0.2  &   0.13 & 0.20 & 0.20  & 0.15   \\
$(D/H)_\odot$  &$(2.6 \pm 1.0) \times 10^{-5}$ & $3.4 \times
10^{-5}$&$2.6 \times 10^{-5}$ &0.0&
         $1.9 \times 10^{-5}$ \\
$(D_p/D)_\odot$ & $\sim$ 3&2.2&3&         &   4     \\
$(D_p/D)_o$  & $\sim$ 5&3.8&5& & 5  \\
$\left(\frac{(D+~^3He)}{H}\right)_\odot$  &$(4.1 \pm 1.0) \times
 10^{-5}$ &$5.1 \times 10^{-5}$&
  $5 \times 10^{-5}$&  $4 \times 10^{-5}$& $3.8 \times 10^{-5}$  \\
$O/O_\odot$ &1&2.0&2.3&1.1&2.6   \\
$Fe/Fe_\odot$ &1&0.9&1.0&0.8&1.1   \\
$Z/Z_\odot$ &1&1.4&1.7&1.3&1.8 \\ \hline
\end{tabular}

\end{center}
\end{table}

\newpage
\beginapjbib
%\bibitem Abia, C., \& Canal, R. 1988, A \& A, 189, 55

%\bibitem Anders, E., \& Grevesse, N. 1989, Geochim. Cosmochim. Acta, 53, 197

\bibitem Arnett, W.D. 1978, ApJ, 219, 1008

\bibitem Arnett, W.D. 1991, in {\it Fontiers of Stellar
Evolution} ed. D, Lambert,
(Mac Donald Observatory), p. 389

\bibitem Audouze, J., Lequeux, J., Reeves, H., \& Vigroux, L.
 1976, ApJ, 208, L51

\bibitem Audouze, J, \& Tinsley, B. M. 1974, ApJ, 192, 487

\bibitem Bania, T. M., Rood, R. T., \& Wilson, T. L. 1993, in {\it Origin
 and Evolution of the Elements} eds. N. Prantzos, E. Vangioni-Flam,
and M. Cass\'{e}
(Cambridge:Cambridge University Press), p. 107

%\bibitem Cameron, A.G.W. 1982, in {\it Essays in Nuclear Astrophysics}, ed. C.
%Barnes, D. Clayton, and D. Schramm (Cambridge:Cambridge University Press),
%%p.35

%\bibitem Cesarsky, C. 1980, ARA\&A, 18, 289

\bibitem Clayton, D.D. 1985, ApJ, 290, 428

\bibitem Coplan, M. A., Ogilvie, K. W., Bochsler, P., \& Geiss, J. 1984,
Solar Phys, 93, 415

\bibitem  Dearborn, D. S. P., Schramm, D., \& Steigman, G. 1986, ApJ, 302, 35

%\bibitem Dearborn, D. S. P., Schramm, D., Steigman, G. \& Truran, J.
%1989 ApJ, 347, 455

\bibitem Delbourgo-Salvador, P., Gry, C., Malinie, G., \& Audouze, J. 1985,
A\&A, 150, 53

\bibitem Delbourgo-Salvador, P. \& Vangioni-Flam, E. 1993,in {\it Origin
 and Evolution of the Elements} eds. N. Prantzos, E.
Vangioni-Flam, and M. Cass\'{e}
(Cambridge:Cambridge University Press), p. 253

\bibitem Deliyannis, C. P., Demarque, P., \& Kawaler, S. D. 1990 ApJS, 73, 21

%\bibitem Deliyannis, C.P.\& Pinsonneault, M.H. 1990 ApJ, 366, L67 (DP)

\bibitem Duncan, D. K., Lambert, D. L., \& Lemke, M. 1992 ApJ, submitted

\bibitem Ferlet,  R. 1992, in {\it Astrochemistry of cosmic phenomena},
Ed. P.D. Singh (Kluwer, Netherlands), p. 85

\bibitem Francois, P., Vangioni-Flam, E., \&  Audouze, J. 1989 , ApJ, 361, 487

\bibitem Geiss, J. 1993, in {\it Origin
 and Evolution of the Elements} eds. N. Prantzos, E.
 Vangioni-Flam, and M. Cass\'{e}
(Cambridge:Cambridge University Press), p. 89

\bibitem Gilmore,G., Edvardsson,B. \& Nissen,
P.E. 1992 Astrophys. J. 378, 17.

\bibitem Gilmore,G., Gustafsson,B., Edvardsson,B. \& Nissen,P.E.
1992, Nature,
357, 379

\bibitem Gott, J. R. III, Gunn, J. E., Schramm, D. N.,
\& Tinsley, B. M. 1974
ApJ, 194, 543

%\bibitem Grevesse, N. \& Anders, E. 1989 in {\sl Cosmic Abundances of Matter},
%ed. C.J. Waddington (AIP Conf. Proc. 183), p. 1.

\bibitem Gry, C., Malinie, G., Audouze, J., \& Vidal-Madjar, A. 1984, in
Formation and Evolution of Galaxies and Large Scale Structure in
the Universe,
eds. J. Audouze \& J. Tran Tranh Van (Reidel, Dordrecht) p 279

\bibitem Hartoog, M.R. 1979, ApJ, 231, 161

\bibitem Hobbs, L.M., \& Duncan, D.K. 1987, ApJ, 317, 796

\bibitem Iben, I. 1967, ApJ, 147, 624

\bibitem Iben, I. \& Truran, J.W. 1978, ApJ, 220,980

\bibitem Iben, I. \& Tutukov, A. 1984, ApJ Supp, 54, 335

\bibitem Kuijken, K., \& Gilmore, G. 1991, ApJ, 367, L9

%\bibitem Kurki-Suonio, H., \& Matzner, R.A., Olive, K.A., \&
%	Schramm, D.N. 1990, ApJ, 353,406

\bibitem Larson, R.B. 1986, MNRAS, 218, 409

\bibitem Linsky, J.L., et al. 1992, ApJ, 402, 694

%\bibitem Malaney, R.A., \& Butler, M.N. 1993 ApJ, 407, L73

%\bibitem Mayle, R.W., Wilson, J.R., \& Schramm, D. N. 1987, ApJ, 318, 288

\bibitem Meneguzzi, M., Audouze, J.,\& Reeves, H. 1971 Astron.
 \& Astrophys., 15,337

%\bibitem Meneguzzi, M., \& Reeves, H. 1975 Astron. \& Astrophys., 40,110

%\bibitem Molaro, P., Castelli, F., \& Pasquini, L. 1993, in {\it Origin
% and Evolution of the Light Elements} eds. N. Prantzos, E. Vangioni-Flam, and
%%M. Cass\'{e}
%(Cambridge:Cambridge University Press), p. 153

\bibitem Norris, J. \& Ryan, S. 1991, ApJ, 380, 403

\bibitem Olive, K.A. 1986, ApJ, 309, 210

\bibitem Olive, K.A., Prantzos, N., Scully, S., \& Vangioni-Flam, E. 1993,
University of Minnesota preprint UMN-TH-1203

\bibitem Olive, K.A., Schramm, D.N., Steigman, G., Turner, M.S., \& Yang,
J. 1981, ApJ, 246, 557

\bibitem Olive, K. A., \& Schramm, D. N. 1992 Nature, 360,439

\bibitem Olive, K. A., Steigman, G., \& Walker, T.P. 1991, ApJ, 380, L1

\bibitem Olive, K.A., Theilemann, F.-K., \& Truran, J.W. 1987,
 ApJ, 313, 813 (OTT)

\bibitem Ostriker, J.P., \& Tinsley, B. 1975, ApJ, 201, L51

\bibitem Ostriker, J.P., \& Schramm, D.N. 1993, in preparation

\bibitem Pagel, B. E. J., Simonson, E. A., Terlevich, R. J., \&
Edmunds, M. 1992, MNRAS, 255, 325

\bibitem Pinsonneault, M. H., Deliyannis, C. P., \& Demarque, P.
1992 ApJS, 78, 179

%\bibitem Prantzos, N., \& Cass\'{e}, M. 1993, ApJ, submitted

\bibitem Prantzos, N. 1993, AA, submitted

\bibitem Prantzos, N., Cass\'{e}, M., \& Vangioni-Flam, E.
 1993a ApJ, 403, 630

%\bibitem Prantzos, N., Cass\'{e}, M., \& Vangioni-Flam, E. 1993b in {\it
%%Origin
% and Evolution of the Light Elements} eds. N. Prantzos, E. Vangioni-Flam, and
%%M. Cass\'{e}
%(Cambridge:Cambridge University Press), p. 156

%\bibitem Rana, N. 1991, ARA\&A, 29, 129

\bibitem Rebolo, R., Molaro, P., Abia, C. \& Beckman, J.E. 1988
 Astron. Astrophys. 193, 193

\bibitem Rebolo, R., Molaro, P., \& Beckman, J.E. 1988, A\&A, 192, 192

\bibitem Rebolo, R., Garcia Lopez, R. J., Martin, E. L., Beckman, J. E.,
McKeith, C. D., Webb, J. K., \& Pavlenko, Y. V. 1993 in
{\it Origin
 and Evolution of the Elements} eds. N. Prantzos, E. Vangioni-Flam,
and M. Cass\'{e}
(Cambridge:Cambridge University Press), p. 149

%\bibitem Reeves, H. 1993, in {\it Origin
% and Evolution of the Elements} eds. N. Prantzos, E. Vangioni-Flam, and M.
%%Cass\'{e}
%(Cambridge:Cambridge University Press), p. 168

\bibitem Reeves, H. Fowler, W.A. \& Hoyle, F. 1970 {\sl Nature}, 226, 727

\bibitem Rood, R.T. 1972, ApJ, 177, 681

\bibitem Rood, R.T., Bania, T.M., \& Wilson, T.L. 1992, Nature, 355, 618

\bibitem Rood, R.T., Steigman, G. \& Tinsley, B.M. 1976, ApJ, 207, L57

\bibitem Ryan, S.,  Bessel, M., Sutherland, R., \& Norris, J.
 1990 ApJ, 348, L57

\bibitem Ryan, S., Norris, J., Bessel, M., \& Deliyannis, C.
1992 ApJ, 388, 184

\bibitem Scalo, J.M. 1986, Fund. Cosm. Phys., 11, 1

\bibitem Smith, V. V., Lambert, D. L., \& Nissen, P. E.
1992 ApJ, submitted

\bibitem Solomon,   1993, private communication

\bibitem Spite, F., \& Spite, M. 1982a Nature, 297, 483

\bibitem Spite, F., \& Spite, M. 1982b A\&A, 115, 357

\bibitem Spite, F., \& Spite, M. 1986 A\&A, 163, 140

\bibitem Steigman, G. \& Tosi, M. 1992, ApJ, 401, 150

\bibitem Steigman, G., \& Walker, T. P. 1992 ApJ, 385, L13

\bibitem Steigman, G., Fields, B., Olive, K. A., Schramm, D. N.,
\& Walker, T. P. 1993, ApJLett (in press)

\bibitem Thielemann, F.-K., Nomoto, K., Hashimoto M. 1993 in
{\it Origin
 and Evolution of the Elements} eds. N. Prantzos, E. Vangioni-Flam,
 and M. Cass\'{e}
(Cambridge:Cambridge University Press), p. 297

\bibitem Thomas, D., Schramm, D. N., Olive, K. A., \& Fields, B. D.
1993 ApJ, 406, 569

\bibitem Tinsley, B.M. 1977, ApJ, 216, 548

\bibitem Tinsley, B.M. 1980, Fund. Cosm. Phys., 5,287

\bibitem Tosi, M. 1988, A\&A, 197, 33

\bibitem Truran, J.W. \& Cameron, A.G.W. 1971, Astrophys. \& Spa. Sci.,
14, 179

\bibitem Twarog, B.A. 1980, ApJ, 242, 242

\bibitem Vangioni-Flam, E., \& Audouze, J. 1988,
A\&A, 193, 81 (VFA)

%\bibitem Vangioni-Flam, E.,  Cass\'{e}, M., Audouze, J., \& Oberto, Y. 1990,
%ApJ, 364, 568

\bibitem Vangioni-Flam, E., Prantzos, N., \& Chauveau, S. 1993,
in preparation

\bibitem Vidal-Madjar, A. 1991, Adv Space Res, 11, 97

\bibitem Vidal-Madjar, A. \& Gry, C. 1984, A\&A, 138, 285

%\bibitem Walker, T. P., Steigman, G., Schramm, D. N., Olive, K. A.,
%\& Fields, B. 1993 ApJ, in press

\bibitem Walker, T. P., Steigman, G., Schramm, D. N., Olive, K. A.,
\& Kang, H. 1991 ApJ, 376, 51

\bibitem Weaver, T., \& Woosley S. E., 1993, Phys. Rep., 227, 65

\bibitem Woosley, S.E. 1993, Les Houches summer school on Supernovae
(to be published)

\bibitem Woosley, S.E. \& Weaver, T.A. 1986, in {\it Radiation
 Hyrodynamics in Stars
and Compact Objects} eds. D. Mihalas \& K.H. Winkler (Dordrecht:Reidel)

\bibitem Yang, J., Turner, M.S., Steigman, G., Schramm, D.N.,
\& Olive, K.A. 1984,
ApJ, 281, 493.

\endapjbib

\newpage
\noindent{\bf{Figure Captions}}

\vskip.3truein

\begin{itemize}

\item[]
\begin{enumerate}
\item[]
\begin{enumerate}
\item[{\bf Figure 1:}] a: Impact of the IRA on the evolution of
$D/H$ (1)
 and $(D +~ ^3He)/H$ (2) as a function
of time for the model $I_{a,e}$. The data points correspond to
Eqs. (\ref{dhe3}), (\ref{dsolar})
and (\ref{do}). Dashed curves correspond to calculation made
with the IRA, and
solid curves are calculated without the IRA.

b: As in Fig.~1a, for the model $II_{a,a}$.

\item[{\bf Figure 2:}] a: The evolution of the gas mass
 fraction as a function
of time, with (dashed curve) and without (solid curve) the
 IRA for model $I_a$.

b: As in Fig.~2a, for model $II_{a,a}$.

\item[{\bf Figure 3:}] The abundances of \d (1), and \he3
 (2) as a function of the
gas mass fraction $\sigma$ with (dashed curve) and without
(solid curve) the IRA for model $II_{a,a}$.

\item[{\bf  Figure 4:}] Metallicity distribution in the
 Galactic disk (dashed lines) for three models:
$I_a$, $II_{a,a}$, and $II_b$.  (Model $I_b$ is quite similar to $I_a$.)
The data shown as a histogram (solid line) are taken from
 Norris \& Ryan (1991).

\item[{\bf Figure 5:}] a: The evolution of the $D/H$ ratio
as a function of time for the
different models: Model $I_{a,e}$ with $\Psi(t) = 0.25
\sigma$ and $\phi(M) \propto
M^{-2.7},   0.4 \le (M/M_\odot) \le 100$; Model $I_b$
with $\Psi(t) = 0.25
\sigma$ and the IMF from Tinsley (1980); Model $II_{a,a}$
with $\Psi(t) = e^{-t/\tau},
\tau = 3$Gyr, and the IMF from Scalo (1986),  $0.1 \le (M/M_\odot) \le 20$;
Model $II_b$ with $\Psi(t) = e^{-t/\tau},
\tau = 3$Gyr, and $\phi(M) \propto
M^{-2.7},   0.4 \le (M/M_\odot) \le 100$.

\item[{\bf  Figure 5:}] b: as in 5a for models $II_a$: $II_{a,a}$ with
$\tau = 3$ Gyr; $II_{a,b}$ with
$\tau = 5$ Gyr; $II_{a,c}$ with
$\tau = 10$ Gyr.

\item[{\bf  Figure 5:}] c: The impact of the
 lower limit of the IMF $M_{inf}$ on the evolution
of ($D/H$) using model $I_{a,e}$. The curve labeled 1 (2)
uses $M_{inf} = 0.4M_\odot$ ($M_{inf} = 0.1M_\odot$).

\item[{\bf  Figure 6:}] a: As in Figure 5a, for the
 evolution of $(D+~^3He)/H$.

\item[{\bf  Figure 6:}] b: As in Figure 5b, for the
 evolution of $(D+~^3He)/H$.

\item[{\bf  Figure 6:}] c: As in Figure 5c, for the
 evolution of $(D+~^3He)/H$.

\item[{\bf  Figure 7:}] Evolution of \he3/H for different galactic models

\item[{\bf  Figure 8:}]  The evolution of $(D +~^3He)/H$ as a
function of time
for Model $I_a$ with different sets of values of the $^3He$ survival
fraction, $g_3$  (sets of values refer to stellar masses
 (1, 2, 3) M$_{\odot}$,
respectively:
Model $I_{a,a}$ with $g_3 = (1,0.7,0.7)$;
Model $I_{a,b}$ with $g_3 = (1,0.5,0.5)$;
Model $I_{a,c}$ with $g_3 = (0.7,0.5,0.5)$;
Model $I_{a,d}$ with $g_3 = (0.7,0.3,0.3)$;
Model $I_{a,e}$ with $g_3 = (0.5,0.3,0.3)$.

\end{enumerate}
\end{enumerate}
\end{itemize}

\end{document}